\documentclass[runningheads]{llncs}
\usepackage[T1]{fontenc}
\usepackage{graphicx}
\usepackage{xspace}
\usepackage{subcaption}
\usepackage{hyperref}

\usepackage{comment}

\begin{document}

\newcommand{\infrastructureName}{\textsc{Grid'5000/ABACA}\xspace}

\title{The INRIA DataLake: A Generic and Scalable Ecosystem of Pipelines for HAL Applied to Software Mentions Tracking}

\titlerunning{The INRIA DataLake}

\author{Luca Foppiano\inst{1,3,5}\orcidID{0000-0002-6114-6164}\thanks{Corresponding author: \email{luca@sciencialab.com}} \and
Vipul Gupta\inst{2}\orcidID{0000-0002-6311-4422} \and 
Samuel Scalbert\inst{3}\orcidID{0009-0002-0423-9281} \and
Estelle Nivault\inst{3}\orcidID{0000-0003-0630-5633} \and
Kumar Guha\inst{3}\orcidID{0009-0005-6649-9515} \and 
Yannick Barborini\inst{4}\orcidID{0000-0003-3756-8647} \and 
Alain Monteil\inst{3}\orcidID{0000-0003-3150-4837} \and 
Laurent Romary\inst{3}\orcidID{0000-0002-0756-0508}}

\institute{ScienciaLAB, Portugal\\
\and
Institute of Materials Physics, Helmholtz-Zentrum Hereon, Germany\\
\and
Inria, France\\
\and
CCSD, CNRS, France\\
\and
Common Crawl Foundation, United States\\
}

\authorrunning{Foppiano et al.}

\maketitle

{
  \renewcommand{\thefootnote}{}
  \footnotetext{This preprint has not undergone peer review or any post-submission improvements or corrections. This work has been submitted to TPDL 2026 (Communications in Computer and Information Science, Springer) for consideration.}
}

\begin{abstract}
Research repositories contain a large amount of scientific knowledge, but access to structured articles and specialised information, such as datasets or software metadata, remains limited. In this paper, we present the INRIA DataLake project, which provides an ecosystem of scalable and interconnected pipelines for preparing scientific literature, extracting structured information, and applying specialised treatments. Using a large-scale shared infrastructure, \infrastructureName, we demonstrate our ecosystem through a concrete use case: extracting software mentions from scientific articles deposited daily and visualising them after validation in the HAL research portal. Our results show that the system can efficiently process large volumes of scientific literature while supporting user validation and interoperability with external systems. Designed to grow by integrating additional pipelines and sharing the preparation effort across research groups, this project already contributes to open science through improved visibility and tracking of research software.

\end{abstract}

\section{Introduction}

Digital repositories host a rapidly growing volume of scientific knowledge. However, most of this information remains embedded and dispersed in scholarly articles~\cite{khabsa2014the}, which limits its accessibility and reuse.
Scientific research suffers from a persistent lack of data mutualisation, leading teams to repeatedly duplicate efforts in data preparation.
While traditional search systems support keyword-based retrieval, they do not provide access to structured data or to important research outputs, such as software, datasets, or experimental details. 
In recent years, open science policies have emphasised the need to better capture and measure research production beyond publications. 
For example, in France, the Open Science Monitor~\cite{bassinet2023large} reflects this evolution by aiming to quantify not only citation-based impact, but also the production of software and datasets. 
Achieving this objective requires scalable methods to extract structured information from large document collections.

Data lakes have emerged as a dominant paradigm for storing and processing large volumes of heterogeneous data~\cite{giebler2019leveraging,10.1145/3132847.3133171}. While classical data lake architectures defer transformation to the point of use, scientific literature processing benefits from early structured extraction — converting PDFs into machine-readable representations that downstream pipelines can consume without repeated parsing effort. Existing data lake solutions remain largely general-purpose and do not address these domain-specific needs, such as integration with scholarly repositories and open science infrastructures.

In this paper, we present an ecosystem of generic, scalable, and interconnected pipelines for processing scholarly articles, extracting structured information, and enabling post-hoc analytics and insight generation. The system is designed to work at scale, and to support multiple extraction tasks in an asynchronous way. The pipelines are executed on \infrastructureName~\cite{grid5000,bolze2006grid,badia2013enabling}, a large-scale shared computing infrastructure, enabling processing at the scale required by a national repository.

The ecosystem is built around publications from HAL~\cite{hal}, the French national open repository operated by the CCSD on behalf of CNRS, Inria, and INRAE. HAL centralises more than 150 institutional archives (HAL portals) under a single multidisciplinary database, hosts dedicated spaces for theses (HAL Theses), the humanities and social sciences (HAL-SHS), and multimedia content (MediHAL), and provides a channel for declaring research software in connection with Software Heritage — combining the scale, disciplinary breadth, and institutional anchoring that make it a particularly relevant testbed for this project. 
As a concrete demonstrator, we present the use case involving extraction and validation of software mentions from pre-processed structured scientific articles. 

Our contribution is twofold. First, we introduce a modular data processing workflow adapted to a shared resource environment, that combines document synchronisation, information extraction, and notification mechanisms. 
Second, we demonstrate the concrete deployment of this architecture on the HAL repository through an end-to-end use case — software mention extraction, human validation, and dissemination to external services — confirming its suitability for a national-scale production environment.

\section{System Architecture}

The proposed system is designed as a generic infrastructure for large-scale document processing, where scientific literature is collected, processed, and enriched through a series of independent but coordinated pipelines. 
The architecture accommodates multiple types of extraction tasks — such as software mentions, datasets, or quantitative information — without requiring significant changes to the core design. 

The system runs on \infrastructureName, which provides bare-metal access to around 15,000 cores across 800 compute nodes with diverse hardware profiles, including GPUs, NVMe storage, and high-speed networks. This is particularly suited to our needs, as each pipeline stage has different resource profiles — CPU-bound parsing versus GPU-bound extraction — and running them on shared infrastructure substantially reduces both cost and carbon footprint compared to dedicated cloud deployments. 

\begin{figure}[htbp]
  \centering
  \includegraphics[width=0.95\linewidth]{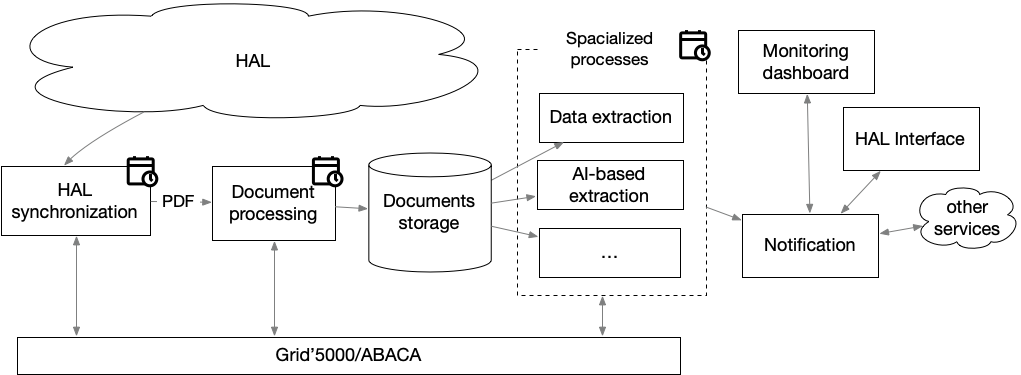}
  \caption{Overview of the system.}
  \label{fig:overview}
\end{figure}

A schematic overview is shown in Fig.~\ref{fig:overview}.

Each pipeline is scheduled independently on \infrastructureName, and runs whenever the requested resources become available. Resources can be specified granularly, from number of hosts down to CPUs, GPUs, and memory per host. The only coupling between stages is the availability of files produced by the preceding one: if data is not available at scheduling time, the pipeline processes it at the next run, ensuring flexibility and avoiding bottlenecks when, for example, GPU nodes are temporarily unavailable. Each pipeline follows the same internal pattern: a preprocessor partitions documents into a staging area, a processor applies the transformation, and a post-processor verifies and retains valid outputs. The staging area isolates the processing environment from the canonical HAL mirror, making failure recovery a matter of re-staging the affected chunk rather than reprocessing the whole corpus. A local database tracks the processing state of each document throughout its lifecycle.

\subsection{HAL Synchronisation}

The first pipeline is responsible for synchronising data from the HAL repository. 
When the job starts, it queries the HAL API to retrieve the list of HAL identifiers that have been added or modified during the past 24 hours. The returned set is compared with the list obtained on the previous day, so that records which have been removed can be detected and removed, together with their associated files, from the local mirror.
 
In parallel, new and modified records are downloaded locally: metadata and the PDF document when available.
Documents under embargo are not downloaded, but kept in a dedicated watch list and are pulled automatically as soon as they become publicly available. The combination of daily differential updates and embargo tracking ensures that the local dataset remains consistent, up to date, and complete.

\subsection{Document Processing}

This pipeline is responsible for transforming raw PDF documents into structured representations. 
Parsing is performed using GROBID~\cite{grobid}, which converts PDF documents into structured XML TEI (Text Encoding Initiative) format. 
PDFs are removed after processing, while the GROBID output in TEI-XML is retained for further processing and made available to any interested party.

\subsection{Specialised processes}

The structured TEI representations produced by the parsing pipeline serve as input to one or more specialised processes. This stage is deliberately generic: any extraction component that consumes TEI and produces an enriched output can be plugged in following the same pattern, simply by instantiating a new staging area and a new state-tracking database. Example processes include dataset detection, quantity and measurement extraction, and experimental data extraction.

\subsection{Notification}

Once processing is completed, the extracted information is collected and loaded into a dedicated application acting as a COAR (Confederation of Open Access Repositories) Notify inbox for our ecosystem. 
The COAR Notify Initiative\footnote{\url{https://coar-notify.net/}} is developing and accelerating community adoption of a standard, interoperable, and decentralised approach to linking research outputs hosted in the distributed network of repositories with resources from external services, such as overlay-journals and open peer review services.
This component manages the postprocess of mentions (e.g., filtering mentions that are obvious mistakes)  as well as outbound communication with external systems, and maintains a graph database of documents and their associated extracted information — distinct from the per-pipeline state-tracking databases described above, which record processing status at the document level. 
The graph database serves as the integration layer, linking documents, extracted entities, and external services.

\subsection{HAL enrichment with optional human validation}

A central aspect of the system is the integration of extracted information back into the HAL infrastructure, where it is made available for review by authorised users. 
Human validation is an optional but encouraged step: the pipeline functions without it, but validation significantly improves data quality by allowing authors and librarians to confirm or reject individual mentions before they are propagated to external services. 

\section{Tracking Software Mentions in HAL}
\label{sec:usecase}

As a concrete demonstrator of the ecosystem, we present the end-to-end pipeline for extracting, disseminating, and validating software mentions from scientific articles available in HAL. This use case was selected because software is a first-class research output that remains largely invisible in traditional bibliometric systems, and because it exercises all components of the ecosystem: document parsing, specialised extraction, notification, and human validation.

\subsection{Software-mention Extraction}

The structured TEI representations produced by the GROBID parsing pipeline serve as input to the software-mention extraction component, developed within the SoftCite project~\cite{du2021softcite}. The software-mention extractor takes TEI or PDF documents as input and produces JSON annotations identifying software mentions along with their textual context. 
Extraction jobs are dispatched as a batch of parallel workers on \infrastructureName with the number of nodes adjusted to the workload at hand.
This elastic modulation of resources is a key factor in keeping both operational cost and energy consumption low. Documents for which no software mention is detected do not trigger any downstream notification, which helps keep the load on external systems manageable.

\subsection{Notification and Dissemination}

Once extraction is complete, the results are forwarded through the notification pipeline using COAR Notify, as described in the previous section, to any subscribing service — in this case HAL and Software Heritage~\cite{di2019how}, a universal software source code archive.
The notification payload includes the information most relevant for downstream processing: the name of the detected software and the list of textual \emph{contexts} — the snippets of the source document in which the mention appears — so that human validators can assess a mention at a glance. The notification layer is handled asynchronously, and the outcome (success or failure) is persisted in the local database so that failed deliveries can be replayed without reprocessing the underlying document.

\subsection{Human Validation}

\begin{figure}[htbp]
  \centering
  \includegraphics[width=0.95\linewidth]{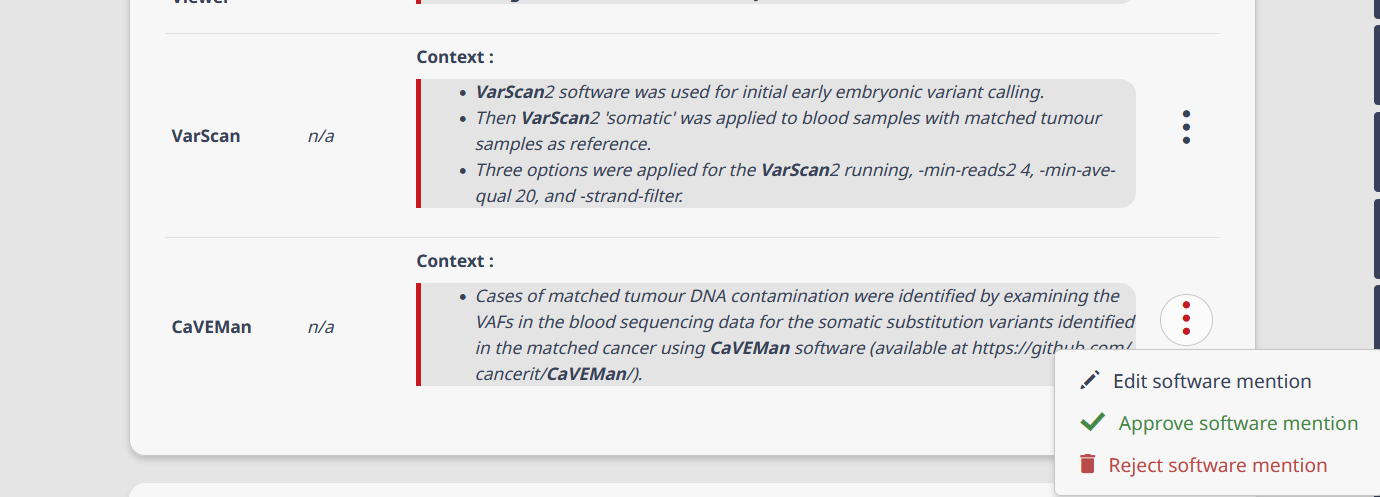}
  \caption{Screenshot of the HAL interface tracking software information in a publication.}
  \label{fig:hal-validation-interface}
\end{figure}

Extracted software mentions are displayed within the HAL interface, where authors, librarians, and other authorised users can review them in context.
A representative example is shown in Fig.~\ref{fig:hal-validation-interface}. Due to repository policies, mentions are only visible to authorised users, such as the article authors or administrators. 
Users can accept or reject each mention individually, and their decisions are sent back to the ecosystem through the same notification mechanism. 
Accepted software mentions are also forwarded to Software Heritage for archival and assignment of a persistent SWHID identifier (ISO/IEC 18670:2025\footnote{\url{https://www.iso.org/standard/89985.html}})~\cite{di2018identifiers}.

\subsection{Evaluation}

The system was evaluated through a large-scale deployment on the HAL repository across two phases. 
In the first phase, the system was tested by processing a large portion of HAL, approximately 1.6 million documents, primarily to validate throughput and identify bottlenecks before moving to production. 
The extracted text from the Grobid XML-TEI output, comprising 1.1 million documents, is available on HuggingFace~\footnote{\url{https://huggingface.co/datasets/lfoppiano/datalake-inria-1.6M-jsonl}}.
The document parsing stage required around one week with GROBID running batches of 100,000 documents over 20 hosts. The software-mention extraction stage took approximately two weeks. 
In the second phase, following the insights gained from the initial tests, the system was deployed in continuous production operation, processing newly deposited documents on a daily basis starting from 1 January 2026. 
After approximately five weeks of continuous operation, around 220,000 documents had been processed end-to-end.

\begin{figure}[htbp]
  \centering
  \includegraphics[width=0.85\linewidth]{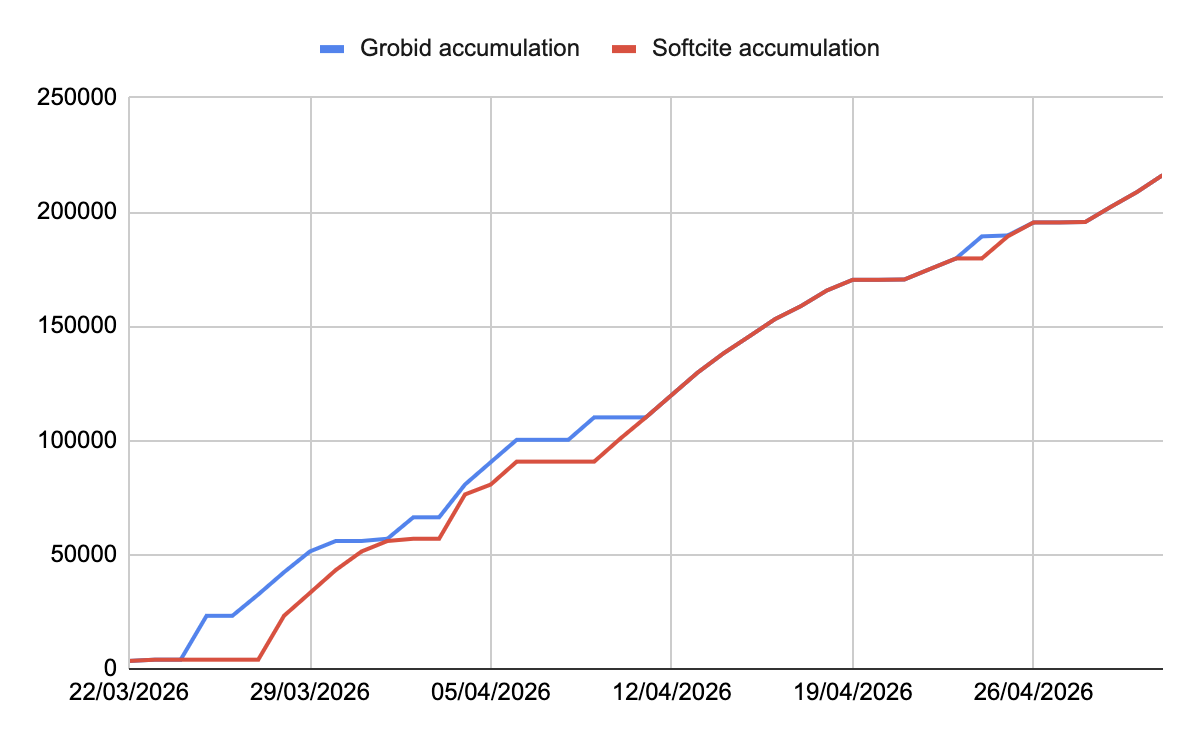}
  \caption{Accumulated document counts over time for the production environment.}
  \label{fig:accumulated-process}
\end{figure}

Fig.~\ref{fig:accumulated-process} illustrates the accumulated document counts over the production period, showing the system recovering from scheduling delays and maintaining overall progress with a fixed allocation of compute hosts.

The present evaluation focuses on throughput and operational robustness. A quality assessment of the extracted software mentions — in terms of precision and recall — is outside the scope of this paper; we refer the reader to the original SoftCite evaluation~\cite{du2021softcite} for model-level performance metrics. Evaluation of the end-to-end pipeline quality, including the impact of GROBID parsing errors on downstream extraction, is left to future work.

% RAW data (DO not share: https://docs.google.com/spreadsheets/d/1KUPeaaCliRLrMH3DCRU_P-03v2kRnqnzsaDq4RKRTog/edit?usp=sharing 

% TODO: Add statistics about the extracted software.
% VG: get subject area from the xml (we can focus on four subject areas (chemistry, computer science, social science, physics) and corresponding mentioning of software. Here, we can restrict to top N software.
% get top N software and plot it against the year to see its use in the science.
% 

\section{Discussion}

\begin{figure}[htbp]
  \centering    
  \begin{subfigure}[b]{0.53\linewidth}
    \includegraphics[width=\linewidth]{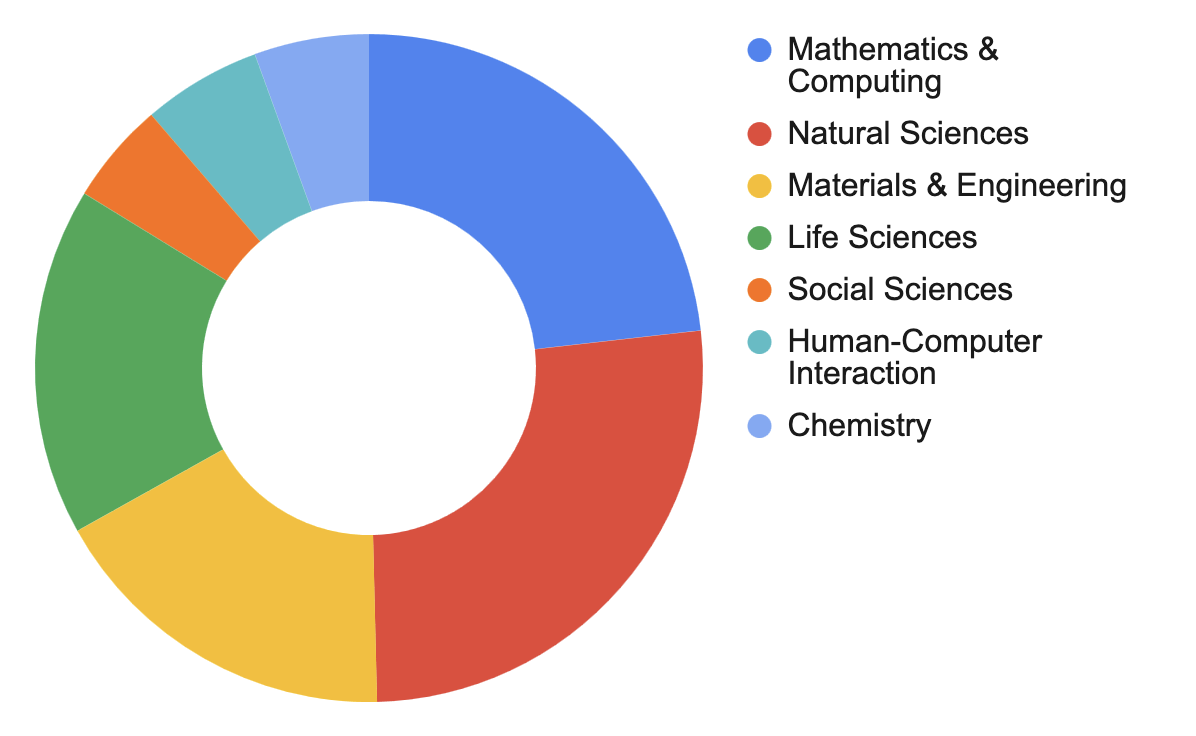}
    \caption{Topic distribution.}
    \label{fig:topic-distribution}
  \end{subfigure}
  \hfill
  \begin{subfigure}[b]{0.44\linewidth}
    \includegraphics[width=\linewidth]{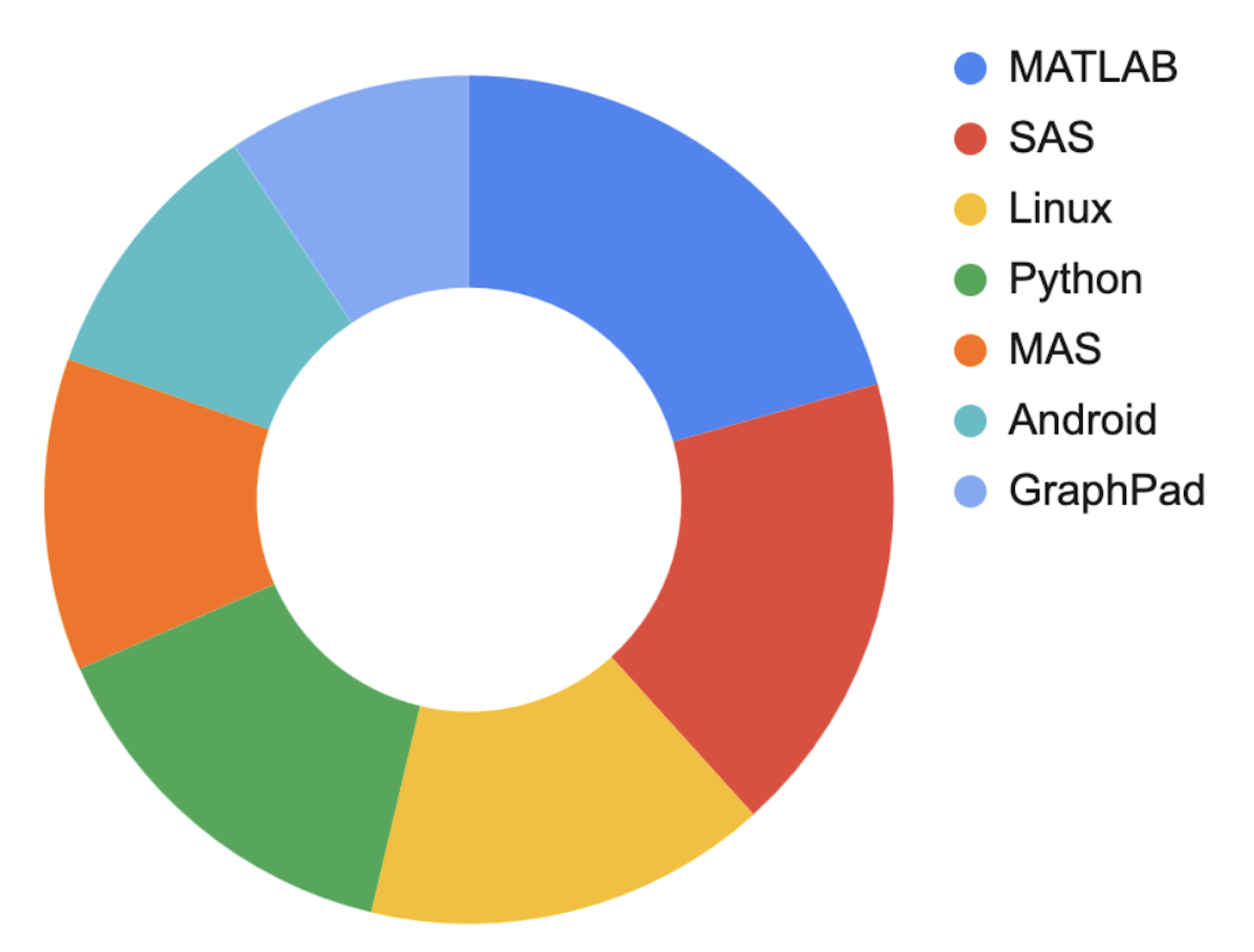}
    \caption{Software distribution. }
    \label{fig:software-distribution}
  \end{subfigure}
  \caption{Topic and software distributions in the 220,000 papers processed in production. Mathematics \& Computing accounts for the largest share of software mentions (65k), dominated by MATLAB, Python, and Linux; Social Sciences follows (20k, mainly SAS, SPSS, Excel); Life Sciences ranks third (19k, with GraphPad, IPA, and ATLAS).}
  \label{fig:distributions}
\end{figure}

The proposed system demonstrates how large-scale document processing can be integrated into existing research infrastructures. 
Software mentions across the corpus (Fig.~\ref{fig:software-distribution}) reflect the thematic distribution of the collection (Fig.~\ref{fig:topic-distribution}), with computational tools dominating in Mathematics \& Computing, statistical packages in Social Sciences, and domain-specific tools in Life Sciences.

By combining automated extraction with user validation, the ecosystem of several interconnected pipelines provides a balanced approach that leverages both computational efficiency and human expertise. 
The modular architecture makes the system adaptable to different use cases. While this paper focuses on software mentions, the system can be applied to other types of information extraction, such as named entities, scientific vocabularies, and citations. This flexibility is essential for supporting evolving open science requirements.
Furthermore, the integration with external services such as Software Heritage highlights the importance of interoperability in modern research ecosystems, and the use of a standardised notification protocol keeps the coupling between repositories, extraction services, and software archives loose enough to evolve independently.

Despite its advantages, the system has several limitations. 
Notifications are not sent for previously published records, which limits the coverage of the system. Extending the current pipelines to historical data requires careful consideration to avoid overwhelming users. 
Finally, user engagement with the validation interface remains an early-stage indicator at the time of writing. Quantitative uptake data — the proportion of mentions reviewed and the accept/reject ratio — will be reported as the system matures and a sufficient validation sample accumulates. More broadly, authors are often reluctant to interact with metadata systems, and measuring and improving engagement with the validation interface is a priority for future work.

\section{Conclusion}

In this paper, we presented an ecosystem of scalable, generic, and interconnected pipelines for processing research documents and extracting structured information. The system was successfully deployed on the HAL repository and demonstrated its ability to handle large-scale workloads in a production setting. 
By focusing on software mentions, we demonstrated how the system can contribute to improving the visibility of research software and support open science initiatives such as the Open Science Monitor and the integration with external services such as Software Heritage. 

Future work will focus on improving user interaction, extending the range of supported extraction tasks, and increasing coverage across historical data.

\begin{credits}
\subsubsection{\ackname} 
The authors would like to thank Patrice Lopez and James Howison, whose work on the SoftCite project provided the dataset and methodology underpinning the software-mention extraction component used in this work.
Part of the work discussed in this article was financed by the SoFAIR project ``Making Software FAIR: A machine-assisted workflow for the research software lifecycle'' (CHIST-ERA-22-ORD-08).

\subsubsection{\discintname}
The authors have no competing interests to declare that are relevant to the content of this article.
\end{credits}

\bibliography{bibliography}
\bibliographystyle{splncs04}

\end{document}